\documentclass[prd,showpcs,amsmath,amssymb,nofootinbib,longbibliography,twocolumn,showpacs,notitlepage,superscriptaddress]{revtex4-1}
\usepackage{multirow}
\usepackage{epsfig}
\usepackage{amsmath}
\usepackage{bm}
\usepackage{times}
\usepackage{graphicx}
\usepackage{color}
\usepackage{slashed}
\usepackage{graphicx}
\usepackage{amsmath} 
\usepackage[latin1]{inputenc}
\usepackage{hyperref}
\usepackage{soul}
\usepackage{multirow}
\usepackage{epsfig}
\usepackage{amsmath}
\usepackage{bm} 
\usepackage{times}
\usepackage{graphicx}
\usepackage{color}
\usepackage{slashed}
\usepackage{graphicx}
\usepackage{amsmath}
\usepackage{tikz}
\usetikzlibrary{positioning,shapes}
\usepackage{relsize} 
\usepackage[latin1]{inputenc}
\usepackage{tikz}
\usetikzlibrary{trees}
\usetikzlibrary{decorations.pathmorphing}
\usetikzlibrary{decorations.markings}
\usetikzlibrary{positioning,arrows}
\usetikzlibrary{decorations.pathmorphing}
\usetikzlibrary{decorations.markings}
\usetikzlibrary{decorations.pathreplacing,calc}
\usetikzlibrary{decorations.pathmorphing,decorations.markings,trees,positioning,arrows}   
\newif\ifmirrorsemicircle
\usetikzlibrary{decorations.pathreplacing,decorations.markings,arrows}

\pgfarrowsdeclarecombine[-5pt]{circled}{circled}{latex}{latex}{o}{o}
\pgfarrowsdeclaredouble{doubled}{doubled}{stealth}{stealth}

\def\bea{\begin{eqnarray}}
\def\eea{\end{eqnarray}}
\def\bean{\begin{eqnarray*}}
\def\eean{\end{eqnarray*}} 
\def\nn{\nonumber}
\def\beaal{\begin{align}}
\def\eeaal{\end{align}}

\begin{document} 
 
\title{Gravitational Waves from Mini-Split SUSY}

\author{Bartosz~Fornal}
\affiliation{Department of Physical Sciences, Barry University, Miami Shores, Florida 33161, USA\vspace{0.8mm}}
\author{Barmak~Shams~Es~Haghi}
\affiliation{Department of Physics and Astronomy, University of Utah, Salt Lake City, Utah 84112, USA\vspace{0.8mm}}
\author{Jiang-Hao~Yu}
\affiliation{CAS Key Laboratory of Theoretical Physics, Institute of Theoretical Physics,\\
Chinese Academy of Sciences, Beijing 100190, China\vspace{0.8mm}}
\affiliation{School of Physical Sciences, University of Chinese Academy of Sciences, Beijing 100049, China\vspace{0.8mm}}
\affiliation{Center for High Energy Physics, Peking University, Beijing 100871, China\vspace{0.8mm}}
\affiliation{School of Fundamental Physics and Mathematical Sciences, Hangzhou Institute for Advanced Study, UCAS, Hangzhou 310024, China\vspace{0.8mm}}
\affiliation{International Centre for Theoretical Physics Asia-Pacific, Beijing/Hangzhou 100190, China\vspace{3.3mm}}
\author{Yue~Zhao\vspace{1mm}}
\affiliation{Department of Physics and Astronomy, University of Utah, Salt Lake City, Utah 84112, USA\vspace{0.8mm}}

\begin{abstract}
\vspace{-2mm}
We show that color-breaking vacua may develop at high temperature in the Mini-Split  Supersymmetry (SUSY) scenario. This can lead to a nontrivial cosmological history of the Universe, including strong first order phase transitions and domain wall production.  Given the typical PeV energy scale associated with Mini-Split SUSY models, a stochastic gravitational wave background at frequencies around 1  kHz is expected. We study the potential for detection of such a signal in  future gravitational wave experiments.  
\vspace{10mm}
\end{abstract}

\maketitle

\section{Introduction}

The direct detection of propagating gravitational waves (GWs) by LIGO \cite{TheLIGOScientific:2014jea} was certainly a milestone discovery. It gave rise to a completely new field, GW astronomy, which is  of great importance   not only for astrophysics, but also high energy physics.  One example of how particle physics  benefits from this great progress, are the opportunities arising from searches for  strong first order phase transitions (FOPTs), cosmic strings, and domain walls (DWs) in the early Universe, directly related to the physics at the high scale. A particularly interesting case   is when the stochastic GW background is produced by a FOPT at the  $\mathcal{O}({\rm PeV})$ scale, since the corresponding signal lies within the sensitivity range of current and future ground-based GW detectors, such as LIGO, Einstein Telescope \cite{Punturo:2010zz} and Cosmic Explorer  \cite{Reitze:2019iox}. 

\vspace{1mm}

To this day, supersymmetry (SUSY) is one of the most appealing frameworks for physics beyond the Standard Model (SM) and predicts the existence of superpartners of SM particles. The SUSY solution to the gauge hierarchy problem suggests  that  masses of superparticles should  not be too far from the electroweak scale. This is the reason why the Minimal Supersymmetric Standard Model (MSSM) \cite{Dimopoulos:1981zb,Dimopoulos:1981yj,Ibanez:1981yh,Marciano:1981un,AlvarezGaume:1981wy}, with superparticles at the $\mathcal{O}(\rm TeV)$ scale, has been thoroughly explored, both experimentally and theoretically. 
In the  MSSM, the charges of superparticles are determined by the quantum numbers of the SM particle content. This  imposes strict constraints  on the shape of the scalar potential. 
Detailed studies of  the zero temperature and finite temperature vacuum structure of the MSSM with ${\rm TeV}$-scale SUSY were carried out in \cite{Carena:1996wj,Carena:1997ki,deCarlos:1997tma,Cline:1999wi,Carena:2008vj,Blinov:2013fta,Hollik:2016dcm}.
However, null results in various SUSY searches at the Large Hadron Collider (LHC) imply that  the  superparticle masses may actually be at higher energy scales. 

\vspace{1mm}

Apart from the direct experimental searches indicating that superparticles should be rather heavy, there are also theoretical arguments in support of this scenario. A generic choice of parameters in the MSSM suffers from flavor-changing neutral current (FCNC) and charge-parity violation  problems unless the masses of superparticles  are at the $\mathcal{O}(100 \ {\rm TeV})$ scale or beyond. In addition, within the MSSM framework, superparticles with  large masses can  increase the Higgs mass, through quantum corrections, from below the $Z$ boson mass up to $\sim 125 \ {\rm GeV}$, so that it agrees with the value  measured at the LHC. A framework with such heavy superparticles is offered by the Mini-Split SUSY \cite{ArkaniHamed:2004fb,ArkaniHamed:2004yi,Arvanitaki:2012ps}. 

\vspace{1mm}

A promising place to look for such high-scale new physics is the early Universe. At temperatures above  $\mathcal{O}(100 \ {\rm TeV})$ the scalar fields corresponding to superparticles in the Mini-Split SUSY become energetically accessible, and their landscape  is highly nontrivial. 
A natural question  is whether the existence of these superparticles generically triggers new phenomena in the history of the early Universe that  can lead to interesting signatures. For example, if the FOPT  happened at temperatures $\mathcal{O}({\rm PeV})$,  the  GW signal would fall within the frequency range accessible by ground-based GW detectors, making them  very powerful probes of high energy physics  which is not accessible by any other terrestrial experiments, e.g., accelerators like the LHC.

\vspace{1mm}

In this study, we demonstrate that a particular realization of the superparticle mass spectrum can lead to an intriguing  evolution of the MSSM scalar potential. Specifically: $(1)$ at a very high temperature, the global minimum of the scalar potential is situated at the origin of the squark field space; $(2)$ as the temperature decreases, one (or several)  of the scalar field directions develops a symmetry breaking minimum away from the origin and a FOPT to the new vacuum may happen; $(3)$ when the temperature drops further,   the symmetry breaking minimum along the squark direction disappears, and the Universe settles  into the electroweak symmetry-breaking minimum along the Higgs direction.
We show   that such an evolution of the Universe leads to the production of GWs, and we discuss its accessibility in future GW detectors.

\section{Symmetry Nonrestoration}

The standard lore is that increasing temperature leads to an enhancement of symmetry. Indeed, for majority of particle physics models, finite temperature effects tend to wash out any nontrivial vacuum structures existing at zero temperature. However, as it was pointed out a long time ago \cite{Weinberg:1974hy}, there are cases  when increasing temperature  actually leads to  a broken phase. Since then, many models of this type have been proposed (see, e.g., \cite{Linde:1976kh,Mohapatra:1979vr,Kuzmin:1981bc,KUZMIN198229,Salomonson:1984rh,FUJIMOTO1985260,Dvali:1995cj,Dvali:1995cc,Lee:1995fb,Rius:1997iy,Bajc:1999cn,Espinosa:2004pn,Sakamoto:2009hb,Meade:2018saz,Baldes:2018nel,Glioti:2018roy,Angelescu:2018dkk,Matsedonskyi:2020mlz,Bajc:2020yvd} and references therein).

An example of such a theory is given by the following extension of the SM Lagrangian, obtained by adding one or more scalar fields $s$,
\bea
 - \mathcal{L} \,=\, - \mathcal{L}_{\rm SM} +  \frac{ \mu_s^2}{2} s^2  + \frac{ \lambda_s}4 s^4 + \frac{\lambda_{hs}}{4}h^2 s^2 \ .
\eea
The mixed quartic coupling $\lambda_{hs}$ can be negative, as long as $\lambda_{hs} \geq -\sqrt{\lambda_s\lambda}$, where $\lambda$ is the Higgs quartic coupling, so that a negative runaway direction is avoided. In the high temperature limit, the thermal Higgs mass squared is given by
\bea
m_h^2(T) \simeq \frac12\left(\frac{\lambda_t^2}{2} + \lambda + N_s\frac{\lambda_{hs}}{6}\right)T^2 \ ,
\eea
where $N_s$ is the number of fields $s$ added into the theory, $\lambda_t$ is the top Yukawa coupling, and we ignored the subleading terms proportional to the electroweak couplings. 
For a  sufficiently negative value of $N_s \lambda_{hs}$, the Higgs thermal mass becomes negative, which  leads to the development of new minima of the effective potential \cite{Meade:2018saz}. 
Such symmetry nonrestoration and formation of new vacuum states 
may result in a FOPT  at high temperatures, which, in turn, would give rise to GW signals.

It has been argued that symmetry nonrestoration at high temperature cannot happen in supersymmetric theories  \cite{Haber:1982nb,Mangano:1984dq,Bajc:1996kj}, with the exception of models with flat directions \cite{Dvali:1998ct,Bajc:1998jr} or a nonzero background charge \cite{Riotto:1997tf}. Although these arguments hold for the MSSM at temperatures above  the mass scale of all superparticles, they do not apply when the temperature is at intermediate scales. In particular,  if there exists  a mass separation among the superparticles, which can be realized naturally in various SUSY breaking scenarios, symmetry nonrestoration  can be achieved, within the framework  of the Mini-Split SUSY, at temperatures lower than the mass scale of the heaviest superparticles.

%\section{MSSM}
%The fields we have are: 6 left-handed (LH) quarks, 6 right-handed (RH) quarks, 6 LH squarks, 6 RH  squarks, 8 gluons, 8 gluinos, 6 LH leptons, 6 LH sleptons, 3 RH leptons, 3 RH sleptons, 4 EW gauge bosons and 4 EW gauginos. 

To show  this explicitly, 
let us focus on the ${\rm SU}(3)_c$ $D$-term contribution to the MSSM scalar potential \cite{Hollik:2018wrr}, 
\bea
V_D= \frac{g_s^2}6  \Bigg(\sum_{\tilde{q}_L} |\tilde{q}_{L}|^2 - \sum_{\tilde{q}_R} |\tilde{q}_{R}|^2\Bigg)^2 ,
\eea
%where the first summation runs over the color index $\alpha$, 
where the sum is over the left-handed (LH) and right-handed (RH) squark flavors $\tilde{q}_L$ and $\tilde{q}_R$. If the squark masses are above the temperature scale, the corresponding thermal corrections to the  effective potential are  suppressed. %Nevertheless, for consistency, we include contributions from all of the squark fields in our calculation.

The mass spectrum for superparticles is model-dependent. Their interactions, especially the charge assignment under the SM gauge group, may affect the masses. For example, in the framework of gauge mediation, the soft SUSY breaking masses of superparticles are generated by gauge couplings through a spurious SUSY breaking sector. We will assume that this or an analogous mechanism makes all LH squarks lighter than the RH squarks.

Furthermore, it is natural to expect that a moderate mass gap appears for squarks of different flavors, similarly to the mass hierarchy for the SM quarks. This can  be induced either by a  flavor dependence in the SUSY breaking mediation mechanism, e.g., when the Yukawa couplings are involved, or through renormalization group  running from a higher energy scale. Motivated by this, we take the soft mass of   $\tilde{d}_{R}$ to be smaller  than those of  the remaining RH squarks. 

With such  assumptions regarding the particle spectrum, at  temperatures above the mass of $\tilde{d}_{R}$ but below the masses of other RH squarks, the $D$-term contribution to the potential  is approximately
\bea\label{treeV}
V_D= \frac{g_s^2}6  \Bigg(\sum_{\tilde{q}_L} |\tilde{q}_{L}|^2 - |\tilde{d}_{R}|^2\Bigg)^2.
\label{potential}
\eea
In the  discussion below, we demonstrate that the effective potential of the MSSM with the tree-level contribution in Eq.\,(\ref{potential}) develops a  symmetry breaking vacuum along the squark $\tilde{d}_{R}$ direction.  Depending on the initial conditions for the evolution of the early Universe, the existence of this vacuum leads to the possibility of a strong FOPT.
%\footnote{We note that a symmetry breaking vacuum develops also in the 1D case, i.e., when only the $\tilde{d}_R$ squark is lighter than the other RH squarks, including $\tilde{s}_R$. This scenario  can lead to a FOPT as well. However, we do not consider this case in more detail, since it does not exhibit as rich of a vacuum structure as the 2D case.}

\section{Thermal effective potential}\label{EffP}

Without loss of generality, we consider the color direction  $(1,0,0)$ and  analyze the potential in terms of the field $\tilde{d}_{R1}$.
Upon separating the  real and imaginary parts,
\bea
\tilde{d}_{R1} &=& \frac{1}{\sqrt2}\left(\phi_d+ i a_d\right) , 
\eea
the  effective potential becomes a function of $\phi_d$
and  consists of three contributions: tree-level  $V_{\rm tree}(\phi_d)$, one-loop Coleman-Weinberg  $V_{\rm {loop}}(\phi_d)$ and finite temperature  $V_{\rm temp}(\phi_d,T)$,
\bea
V_{\rm eff}(\phi_d, T) = V_{\rm tree}+ V_{\rm {loop}}+ V_{\rm temp} \ .
\eea
The relevant contributions to the tree-level part come from the soft masses and the ${\rm SU}(3)_c$ $D$-term in Eq.\,(\ref{treeV}), 
\bea\label{sss}
V_{\rm tree} = \frac12 m_{\tilde{d}}^2 \,\phi_d^2  + \frac{g_s^2}{24}  \,\phi_d^4  \ . 
\eea
Since we are considering the potential in the direction of the RH squark $\tilde{d}_{R1}$, the ${\rm SU}(2)_L$ $D$-term does not contribute to the field-dependent masses. We also  neglect the  ${\rm U}(1)_Y$ $D$-term, since its effect is subleading.
%thus  the quarks, leptons, sleptons, electroweak gauge bosons and electroweak  gauginos do not develop field-dependent masses. 

Gauginos and higgsinos are usually assumed to be lighter than squarks in the Mini-Split scenario, due to $R$-symmetry and Peccei-Quinn symmetry. However, their masses are free parameters and depend on the details of the model. For simplicity, we assume that the gauginos and higgsinos are much heavier, and the corresponding interactions, such as the gaugino-squark-quark coupling, can be neglected. Denoting by  $m_i^2$ the eigenvalues of the 
field-dependent squared mass matrix $m_{i,j}^2$ given by
\bea\label{eee}
m_{i,j}^2(\phi_d) &=& \frac12\frac{\partial^2 V}{\partial [{\rm Re}(\tilde{q}_{i})] \partial [{\rm Re}(\tilde{q}_{j})]} \bigg|_{{\rm Re}(\tilde{d}_{R1})=\frac{\phi_d}{\sqrt2} } \nn\\
&+&  \frac12\frac{\partial^2 V}{\partial [{\rm Im}(\tilde{q}_{i})] \partial [{\rm Im}(\tilde{q}_{j})]} \bigg|_{{\rm Im}(\tilde{q}_{i})=0} \ ,
\eea
we arrive at the squared masses:
$ g_s^2 \phi_d^2/2$ for ${\rm Re}(\tilde{d}_{R1})$,  $  g_s^2 \phi_d^2/6$ for the remaining real and imaginary components of $\tilde{d}_{R}$, and $-  g_s^2 \phi_d^2/6$ for  the LH squark components.

The Coleman-Weinberg contribution is \cite{Quiros:2007zz}
\bea\label{fdm}
V_{\rm loop}&=&\sum_{i}\frac{n_i}{64\pi^2} \bigg\{m_i^4\left[\log\left(\frac{m_i^2}{\mu^2}\right)-c_i \right] \!\bigg\} \ ,
\eea
where the sum is over all particles in the theory with field-dependent masses, $n_i$ is the number of degrees of freedom with an extra minus sign for fermions, $c_i = 3/2$ for scalars and fermions, $c_i = 5/6$ for vector bosons, and $\mu$ is the renormalization scale.

%the Coleman-Weinberg contribution, upon implementing the cutoff regularization scheme and setting  the minimum of the one-loop potential at  its tree-level value \cite{PhysRevD.45.2685,Quiros:2007zz}, is given by
%\bea\label{fdm}
%V_{\rm loop}&=&\sum_{i}\frac{n_i}{64\pi^2} \bigg\{m_i^4(\phi_d, \phi_s)\left[\log\left(\frac{m_i^2(\phi_d, \phi_s)}{M_i^2}\right)-\frac32 \right]\nn\\
%&+& \ 2 m_i^2(\phi_d, \phi_s)\,M_i^2 \bigg\}\ , \ \ \ \ \ 
%\eea
%where the sum is over all the particles in the theory that have nonzero field-dependent masses, $n_i$ is the number of degrees of freedom with an extra minus sign for fermions, and $M_i$ are the soft masses.

The temperature-dependent part of the effective potential  consists of  the one-loop finite temperature term and, in case of bosons, the contribution from Daisy diagrams,
\bea
V_{\rm temp} = V^{(1)}_{\rm temp} + V^{(2)}_{\rm temp} \ ,
\eea
where \cite{Quiros:2007zz}
\bea\label{FT}
V^{(1)}_{\rm temp} &=& \frac{T^4}{2\pi^2} \sum_{i}  n_i \int_0^\infty \!dy\, y^2  \log\left(1\mp e^{-\sqrt{{m_i^2}/{T^2}+y^2}}\right) ,\nn\\
V^{(2)}_{\rm temp}&=& -\frac{T}{12\pi} \sum_{j} n'_j \left\{ \left[m_j^2+\Pi_j(T)\right]^{\frac32}-m_j^3\right\}  \, ,
\eea
with the  negative  sign for bosons and the positive  sign for fermions. The sum over $i$  includes all particles with field-dependent masses, and the sum over $j$ includes only bosons. 
The squark and gluon  thermal masses $\Pi_j(T)$ can be  calculated following the prescription  in \cite{Comelli:1996vm,Huang:2020bbe} and are given by
\bea\label{eq12}
\Pi_{j}(T)  = \frac{4}{9}g_s^2T^2  \ , \ \ \ \  \Pi_g(T) =\frac{19}6 g_s^2 T^2  \ .
\eea

We note that some field-dependent squared masses are negative, resulting in the one-loop zero temperature and finite temperature contributions to the potential developing imaginary parts. However, as  shown in \cite{Delaunay:2007wb} (see also \cite{Curtin:2016urg}), the imaginary terms cancel between the various contributions (this happens for any $T$, not only at high temperature), and the total effective potential is  real. 

\begin{figure}[t!]
\includegraphics[trim={0mm 0mm 0cm 0},clip,width=7.7cm]{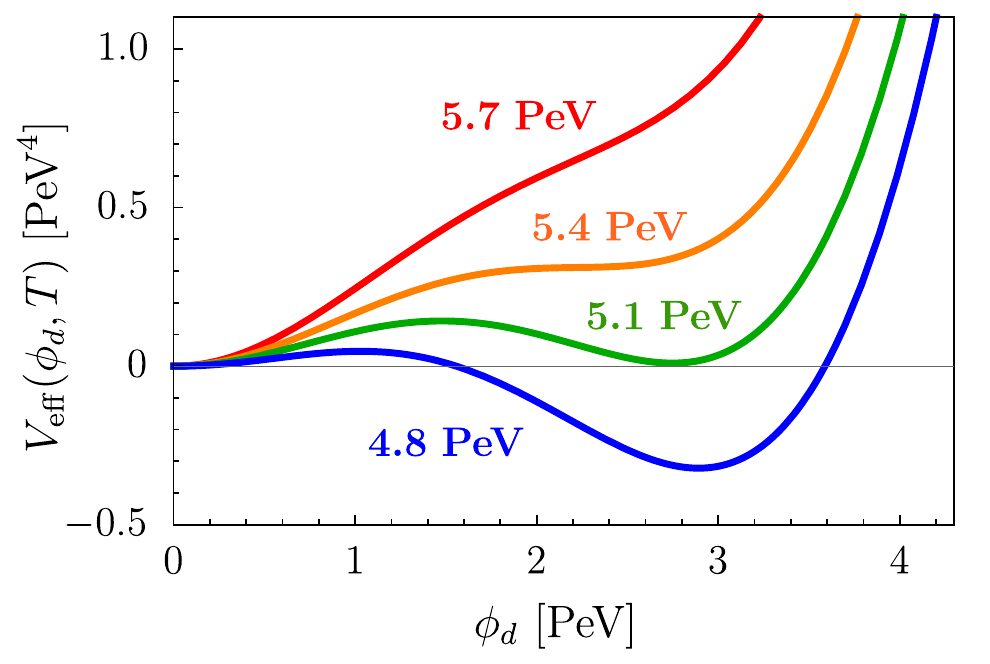}
\vspace{-2mm}
\caption{Effective potential $V_{\rm eff}(\phi_d, T)$ for four temperatures: $4.8 \ {\rm PeV}$, $5.1 \ {\rm PeV}$, $5.4 \ {\rm PeV}$ and $5.7 \ {\rm PeV}$, assuming $m_{\tilde{d}}=1 \ {\rm PeV}$,  $M_{\tilde{q}}=10 \ {\rm PeV}$ and upon subtracting off $V_{\rm eff}(0,T)$. The nontrivial vacuum structure disappears for $T < 1.8 \ {\rm PeV}$.\\}\label{fig1}
\end{figure} 

\section{Vacua in the effective potential}\label{EffP}

Useful intuition can be developed by analyzing the effective potential in the limit 
\bea\label{regg}
m_{\tilde{d}} \ll T \ll M_{\tilde{q}} \ , 
\eea
where $m_{\tilde{d}}$ denotes the soft mass of $\tilde{d}_R$  (assumed to be small), whereas $M_{\tilde{q}}$ is the soft mass scale for the remaining heavy RH squarks.  Denoting collectively $\Pi_j(T) = c_j T^2$,
the finite temperature terms can be approximated by
\bea\label{ewe}
V^{(1)}_{\rm temp} &\simeq& -\frac{\pi^2}{90}T^4 + \frac{T^2}{24} \sum_i n_i\,m^2_i \ , \nn\\
V^{(2)}_{\rm temp} &\simeq& -\frac{T^4}{12\pi}\sum_i  {c_i}^{3/2} - \frac{T^2}{8\pi}\sum_j n_j\,\sqrt{c_j}\, m_j^2 \ .
\eea
Using Eq.\,(\ref{eee}), one arrives at
\bea\label{12}
\sum_j n_j \,m^2_{j} = -\frac{14}{3} g_s^2 \phi_d^2 \ .
\eea
The only other particles developing nonzero  field-dependent masses are five of the  gluons, each with $n_g =3$, yielding
\bea\label{16}
\sum_g n_g m^2_g = 4 g_s^2 \phi_d^2 \ .
\eea
Equations (\ref{eq12}), (\ref{ewe}), (\ref{12}) and (\ref{16}) give the  thermal field-dependent contribution to the effective potential equal to
\bea\label{177}
\Delta V_{\rm eff}  = C \, \phi_d^2\,T^2 \ , \ \ \ 
\eea
where the constant $C$, upon setting $g_s(1 \ \rm PeV)\approx 0.8$, is
\bea
C \approx -0.003 \ .
\eea
Since $C<0$, the effective potential receives a negative contribution to the quadratic terms at finite temperature, which leads to the development of  a vacuum expectation value (vev) with the size governed by the temperature 
\bea
v(T) \sim {T} \ .
\eea
Therefore, in the regime $m_{\tilde{d}} \lesssim T \lesssim M_{\tilde{q}}$, finite temperature effects give rise to symmetry nonrestoration within the framework of the minimal Mini-Split SUSY itself. 
\newpage

Figure \ref{fig1} shows  the  effective potential for several values of the temperature, assuming that the scale of the heavy RH squarks is $M_{\tilde{q}}= 10 \ {\rm PeV}$,  also taken to be the renormalization scale for the Coleman-Weinberg term. The soft mass for the light RH squark $\tilde{d}_R$ is $m_{\tilde{d}} = 1 \ {\rm PeV}$. For temperatures $T < 1.8 \ {\rm PeV}$ the nontrivial vacuum structure vanishes due to the small size of the thermal contributions in Eq.\,(\ref{177}), leaving only one vacuum at $\phi_d = 0$. The symmetry is also restored at 
temperatures $T \gg M_{\tilde{q}}$, thus the scenario is consistent  with the conclusions of \cite{Haber:1982nb,Mangano:1984dq,Bajc:1996kj}.

\section{Gravitational waves}\label{five}

In this section, we investigate the strength and spectrum of GWs produced by FOPTs  in the Mini-Split SUSY. In our calculation we use the exact  formulas for the finite temperature contribution in Eq.\,(\ref{FT}), not just the high temperature expansion of Eq.\,(\ref{ewe}).

If the  reheating temperature $T_R$ of the Universe  is  above the soft masses of all the squarks, a FOPT may take place. In this scenario, at high temperature the effective potential  has only one minimum at $\phi_d =0$, and the field $\tilde d_R$  starts off its post-reheating evolution at the origin. As the temperature decreases, a nontrivial vacuum structure emerges; the effective potential develops a new minimum at $\phi_d  \sim T$, and the Universe can undergo a FOPT from $\phi_d=0$ to this newly developed vacuum. 
Bubbles of true vacuum are nucleated,  expand and eventually fill out the entire Universe. During this process, GWs are emitted from sound shock waves, bubble collisions and magnetohydrodynamic turbulence in the plasma. The frequency of emitted GWs is determined by the scale $M_{\tilde{q}}$. When the temperature drops below the squark masses, the vacuum situated away from the origin disappears, and the Universe  evolves to the electroweak vacuum.

The resulting GW spectrum depends on the shape of the effective potential. It is determined by the Euclidean action for the saddle point configuration interpolating between the false and true vacuum \cite{Linde:1981zj}. The spectrum is described by four parameters: the bubble wall velocity $v_w$ (we assume $v_w =  0.6\,c$),  the nucleation temperature $T_*$,   the strength of the FOPT  $\alpha$, and the duration of the FOPT   $1/\tilde\beta$. 
Upon finding the Euclidean action corresponding to the nucleation temperature, we determined  the GW parameters  $\alpha$ and  $\tilde\beta$  using standard formulas \cite{Ellis:2018mja}. 
The dominant contribution to the GW signal comes from sound waves and is given by\footnote{It has recently been argued that the amplitude of the GW signal from sound waves includes an additional suppression factor \cite{Guo:2020grp}. However, given the large uncertainties in FOPT calculations \cite{Croon:2020cgk,Guo:2021qcq}, Eq.\,(\ref{sound}) serves as an accurate  order of magnitude estimate of the expected signal strength.} \cite{Hindmarsh:2013xza,Caprini:2015zlo}
\bea\label{sound}
 h^2\Omega_s(\nu) \ &\approx& \  (1.86\times 10^{-5})\  \frac{v_w}{\tilde\beta}\left(\frac{\kappa_s\, \alpha}{1+\alpha}\right)^2\left(\frac{100}{g_*}\right)^{\frac13} \ \ \ \ \ \ \ \ \  \nn\\
 &&\  \times \ \frac{\big(\frac{\nu}{\nu_s}\big)^{3}}{\big[1+0.75\,\big(\frac{\nu}{\nu_s}\big)^{2}\,\big]^{\frac72}} \ ,
\eea
where the parameter $\kappa_s$ is \cite{Espinosa:2010hh}
\bea
\kappa_s\approx \frac{6.9\,v_w^{6/5}\alpha}{1.36-0.037\sqrt\alpha+\alpha} 
\eea
and the peak frequency is
\bea\label{sound2}
\nu_s &=& (0.19 \ {\rm Hz} )\, \frac{\tilde\beta}{v_w}\,\left(\frac{g_*}{100}\right)^\frac16\left(\frac{T_*}{1 \ {\rm PeV}}\right) \ .
\eea

\begin{figure}[t!]
\includegraphics[trim={2mm 3mm 0cm 0},clip,width=8.95cm]{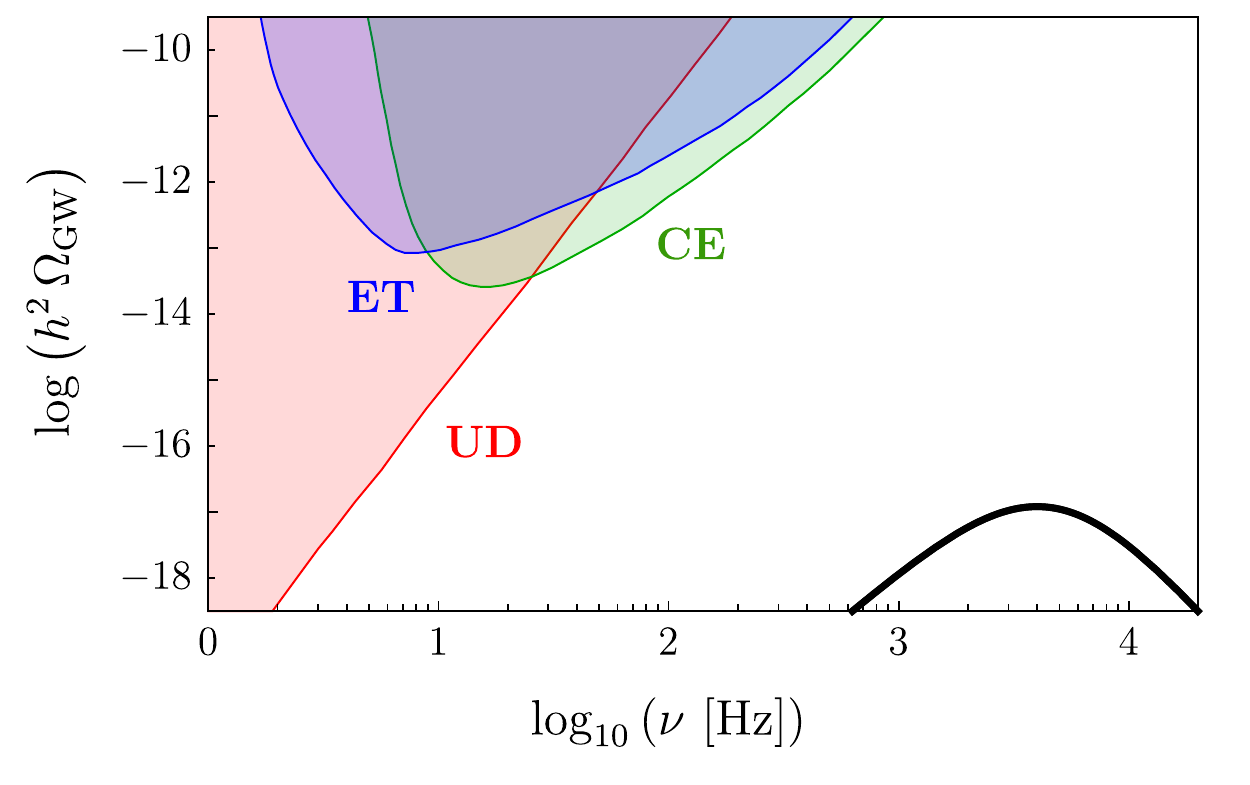}
\caption{Gravitational wave signature (black line) resulting from a first order phase transition in the Mini-Split SUSY  for  $T_R > M_{\tilde{q}} = 10 \ {\rm PeV}$, $m_{\tilde{d}}=1 \ {\rm PeV}$, and  a nucleation temperature $T_* \approx 4.9 \ {\rm PeV}$. Overplotted  are the predicted sensitivities of the GW detectors:\break Einstein Telescope \cite{Sathyaprakash:2012jk} (blue),  Cosmic Explorer \cite{Reitze:2019iox} (green), and\break U-DECIGO \cite{Kuroyanagi:2014qza} (red).
\vspace{0mm}}\label{fig2}
\end{figure}

Figure \ref{fig2} shows the expected  GW signal for the Mini-Split SUSY scenario considered  (black line). Assuming the soft masses $M_{\tilde{q}} = 10 \ {\rm PeV}$ and $m_{\tilde{q}}=1 \ {\rm PeV}$, the nucleation temperature of the FOPT is $T_* \approx 4.9 \ {\rm PeV}$.
 The signal peaks at frequencies $\sim {\rm kHz}$, but its maximal amplitude is  only  $h^2\Omega_{\rm GW} \sim 10^{-17}$, which is below the sensitivity of the near-future GW detectors:  Cosmic Explorer, Einstein Telescope, and Ultimate DECIGO. This is due to the suppression of the  strength parameter, which in this case can be estimated as  $\alpha \sim  \lambda \,v(T)^4 / \left(\pi^2 g_{\rm eff}T^4/30\right)$, where $\lambda$ is the quartic coupling in Eq.\,(\ref{sss}), leading to a small $\alpha\sim 10^{-4}$.

\section{Beyond the MSSM}\label{sec66}

\begin{figure}[t!]
\includegraphics[trim={3mm 5mm 0cm 0},clip,width=8.9cm]{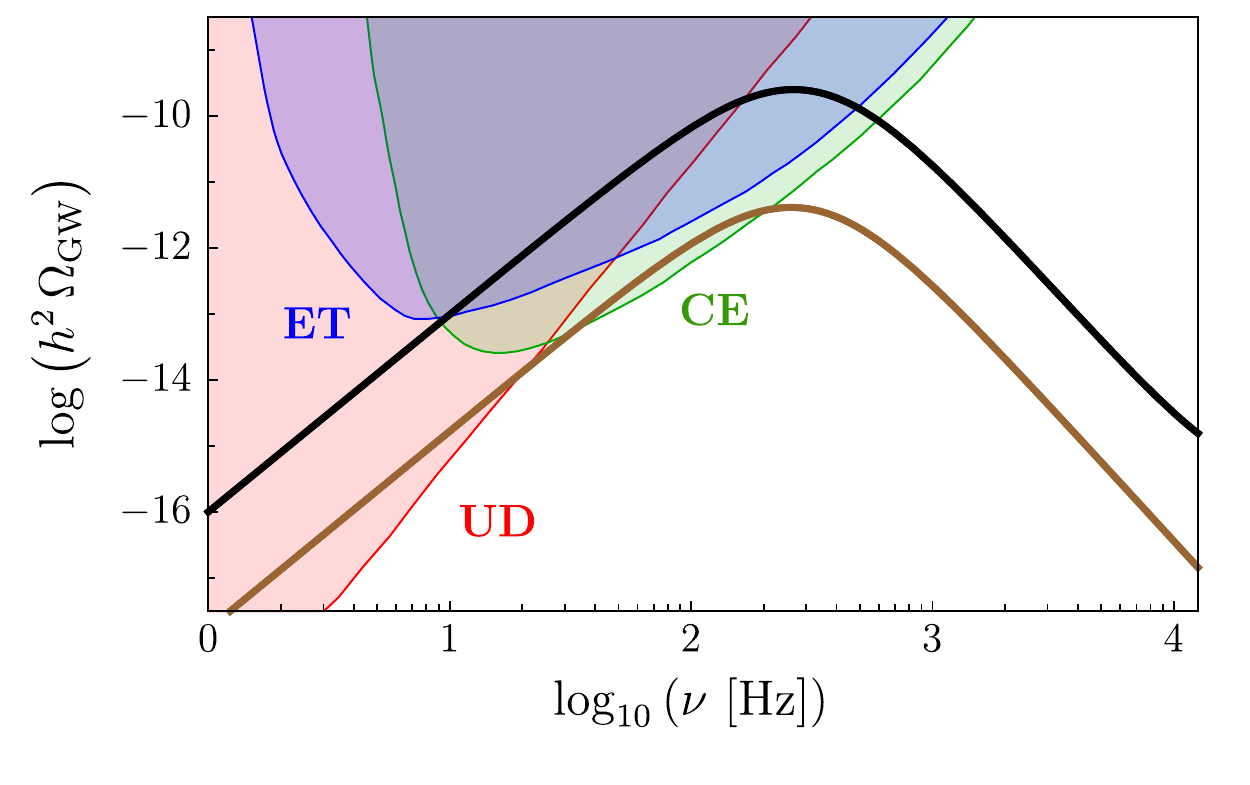}
\caption{Gravitational wave signatures of the MSSM extension with $N=30$ families of squarks. The black curve corresponds to a transition to a one-dimensional vacuum assuming the same parameters as in Fig.\,\ref{fig2}\,; the nucleation temperature is $T_* \approx 1 \ {\rm PeV}$. The brown curve corresponds to a transition between two-dimensional vacua discussed in Sec.\,\ref{sec66} for $M_{\tilde{q}} = 10 \ {\rm PeV}$ and  $ \widetilde{m}_{ds}^2 =0.1 \ {\rm PeV}^2$; the corresponding nucleation temperature is $T_* \approx 600 \ {\rm TeV}$.
Overplotted  are the predicted sensitivities of the GW detectors: Einstein Telescope \cite{Sathyaprakash:2012jk} (blue),  Cosmic Explorer \cite{Reitze:2019iox} (green) and U-DECIGO \cite{Kuroyanagi:2014qza} (red).}
\vspace{6mm}\label{fig3}
\end{figure}

Although the GW signal within the minimal MSSM is beyond the reach of upcoming experiments, it can be significantly enhanced if the MSSM particle spectrum is augmented by a large number of new families of squarks (so that there are $N$ copies of squarks in the theory, where $N\gg 1$). In this case, invoking   a similar mass hierarchy between the squarks as before, i.e., all of the $N$ LH squarks and $\tilde{d}_R$ being much lighter than the remaining $(N-1)$ RH squarks, the resulting  constant in Eq.\,(\ref{177}) becomes  $C \ll -0.003$, leading to a deeper vacuum and an enhanced  GW signal.   Figure\,\ref{fig3} shows the predicted GW signal (black line) for $N=30$ and  the same choice of parameters as in Fig.\,\ref{fig2}\,, which corresponds to a nucleation temperature  $T_* \approx 1 \ {\rm PeV}$.

If two of the RH squarks, e.g., $\tilde{d}_R$ and $\tilde{s}_R$, are lighter than the other RH squarks and both develop vevs, an even richer vacuum structure emerges.
The resulting effective potential exhibits an ${\rm SO}(2)$ symmetry  and has a new minimum  at $\phi_d^2 + \phi_s^2   \sim T^2$, which   constitutes  a  circle in the $(\phi_d,\phi_s)$ field space. Similarly to the previous case, this new vacuum vanishes as the temperature becomes sufficiently small, leaving only one vacuum at $\phi_d=\phi_s=0$.

This setup is sufficient to induce a FOPT at the scale $M_{\tilde{q}}$, which results in GW signals (see, Sec.\,\ref{five}). In addition, some general features which naturally appear in the Mini-Split SUSY scenario may further lead to another FOPT at lower temperatures and the production of DWs. This is because the ${\rm SO}(2)$ degeneracy of the vacuum may be lifted -- extra quartic terms for the squark fields breaking the ${\rm SO}(2)$ symmetry  appear generically upon SUSY breaking. In particular, RH squarks couple to the Higgs field and LH squarks through $A-$terms, e.g., $A_{ab} H_d \widetilde Q_{La} \tilde{d}_{Rb}$ where $a, b$ are flavor indices. Such couplings induce contributions to  $\tilde{d}_{R}$ and $\tilde{s}_{R}$ quartic interactions. The size of such $A-$terms is determined by the SUSY breaking mechanism. In this study, we treat the quartic couplings  in those contributions  as free parameters, 
\bea\label{aaa}
V_{A} =  \left(\lambda_d |\tilde{d}_{R}|^4 + \lambda_s |\tilde{s}_{R}|^4 \right).
\eea
In order to avoid  large thermal corrections to the effective potential induced by those terms, we take $\lambda_d, \lambda_s\ll g_s^2/6$, which makes such contributions much smaller than the ones generated by the $D$-term. The quartic terms in Eq.\,(\ref{aaa}) reduce  the symmetry from ${\rm SO}(2)$ to $Z_4$. This results in a vacuum structure with four  equally deep minima, one  in each quadrant of the $(\phi_d,\phi_s)$ plane. 

With a four-fold degeneracy between the vacua, a FOPT between them would not happen. However, the MSSM offers additional Lagrangian terms breaking this degeneracy. 
One can introduce flavor-changing soft mass terms, which break the $Z_4$ symmetry to $Z_2\times Z_2$. In particular, the MSSM Lagrangian  contains the term
\bea\label{vf}
V_{F} = \widetilde{m}_{ds}^2 \,\tilde{d}_{R}^{\,\dagger} \tilde{s}_{R} + \rm h.c.\ ,
\eea
which breaks the degeneracy between  the vacua and preserves only the $Z_2\times Z_2$ symmetry along two diagonal directions in the $(\phi_d,\phi_s)$ plane.  Although such flavor-changing soft mass terms generically cause phenomenological problems for ${\rm TeV}$-scale SUSY by inducing sizable FCNCs,  those terms are not problematic  in the Mini-Split MSSM  due to  large  squark masses.

Depending on the initial condition and  the reheating temperature, two qualitatively different scenarios occur.  First, the Universe may undergo a FOPT at the heavy RH squark mass scale $\sim M_{\tilde{q}}$, along with a subsequent production of DWs.  Second, the Universe may undergo a FOPT at a scale  $\ll M_{\tilde{q}}$. %These are two qualitatively different scenarios corresponding to two possible choices of the reheating scale:  $(i)$ above or  $(ii)$ below $M_{\tilde{q}}$. 
%\vspace{1mm}

$(i)$ If the reheating temperature of the Universe  is above the soft masses of all the squarks, the effective potential at this temperature has only one minimum at $\phi_d=\phi_s =0$, thus the fields $\tilde d_R$ and $\tilde s_R$ start off their post-reheating evolution at the origin. As the temperature decreases, a nontrivial vacuum structure emerges,  i.e., the effective potential develops a new vacuum at $\phi_d^2 + \phi_s^2 \sim T^2$, as mentioned above. If the $A-$term induced quartic interaction in Eq. (\ref{aaa}) is negligible and the temperature is much higher than the soft mass terms of light squarks, the vacuum is approximately degenerate along a circle. Therefore, the Universe undergoes a FOPT from $\phi_d=\phi_s =0$ to the newly developed vacuum at $\phi_d^2 + \phi_s^2 \sim T^2$. The frequency of emitted GWs is determined by the scale $M_{\tilde{q}}$. With temperature further decreasing,  the degeneracy of the vacuum is broken. Two pairs of vacua are formed with different energy densities  because of  the nonzero flavor-breaking term in Eq.\,(\ref{vf}). This leads to the formation of DWs. The DWs separating nondegenerate vacua annihilate very quickly. On the other hand, the DWs separating degenerate vacua exist  until the temperature drops below the squark masses, i.e., until the vacua situated away from the origin disappear.  After that, the Universe  evolves to the electroweak vacuum. 
%\vspace{1mm}

$(ii)$ The other case is when the reheating temperature is  below the mass scale of the heavy RH squarks $M_{\tilde{q}}$. The squark fields $\tilde d_R$ and $\tilde s_R$ start off with a universal misalignment as the initial condition. At the beginning of the post-reheating era the Universe may reside  in the same  metastable vacuum in all Hubble patches, and, with decreasing  temperature, undergo a strong FOPT to the  true vacuum. The frequency of the produced GWs depends on the scale of $\widetilde{m}_{ds}^2$ and the $A-$term quartic couplings.

As discussed in Sec.\,\ref{EffP}, when the temperature drops to $\mathcal{O}(\rm PeV)$, the $Z_4$ symmetry of the vacuum   is spontaneously broken down to $Z_2\times Z_2$ via the term in Eq.\,(\ref{vf}), and DWs  form around the boundaries of regions corresponding to different vacua. 
The DW dynamics  is governed by the tension $\sigma \sim v(T)^3$ and  the difference between the true and  false vacuum energy densities   $\Delta\rho_{\rm vac}$. In our case $v(T)\sim T$, therefore $\sigma \sim T^3$. The DWs remain stable as long as the tension force $p_T$ given by
\bea
p_T \sim \frac{\sigma \,T^2}{M_{Pl}} \sim \frac{T^5}{M_{Pl}} \ ,
\eea
where $M_{Pl}$ is the Planck mass, is larger than the volume pressure $p_V$, which reads
\bea
p_V \sim \Delta\rho_{\rm vac} \sim \widetilde{m}_{ds}^2 T^2 \ . 
\eea
Since in our case $\widetilde{m}_{ds} \sim  1 \ {\rm PeV}$, the DWs annihilate immediately after forming. This annihilation is the source of   a stochastic GW background. The peak frequency  of  the  GW signal is \cite{Saikawa:2017hiv}
\bea
\nu^{\rm max}_{\rm DW} \sim  (0.1\ {\rm Hz})\ \frac{T_A}{\rm PeV} \ ,
\eea
whereas the amplitude at this  frequency is
\bea
h^2\Omega^{\rm max}_{\rm DW} \sim 10^{-32}\left(\frac{\sigma^2}{T_A^4\times {\rm PeV}^2}\right) \sim 10^{-32}\left(\frac{T_A}{{\rm PeV}}\right)^2 \! . \ \ \ \ \ \ \ \ 
\eea
Thus, the GW signal from DWs is  too weak  to be detected   in the foreseeable future.

We now turn to the case when the reheating temperature is below the mass scale of the heavy RH squarks, and assume that the Universe began its post-reheating evolution in one of the two metastable vacua generated by the $A$-term quartic couplings and the flavor-breaking term. 
 The  scale of the FOPT  is thus governed by the parameter $\widetilde{m}_{ds}^2$. Within the Mini-Split SUSY framework, we choose $m_{\tilde{q}}^2  \sim   \widetilde{m}_{ds}^2 \sim O(0.1) \ {\rm PeV}^2$, which sets the nucleation temperature at $T_* \sim \mathcal{O}({500 \ \rm TeV})$. The only other free parameters in the effective potential are $\lambda_d$ and $\lambda_s$.
For various choices of these parameters, we 
 calculated the Euclidean action at the nucleation temperature  using the software {\fontfamily{cmtt}\selectfont
anybubble} \cite{Masoumi:2016wot}, and used it to compute the FOPT parameters $\alpha$ and $\tilde{\beta}$, which were then  plugged into Eq.\,(\ref{sound}) to derive the expected GW spectrum.

Figure \ref{fig3} presents the GW signal (brown line) in the Mini-Split SUSY extended to $N=30$ families of squarks within scenario $(ii)$. The $A-$term quartic couplings were set to $\lambda_d = \lambda_s = 0.03$ and the flavor-breaking term, along with the renormalization scale, were chosen to be $ \widetilde{m}_{ds}^2 = \mu^2=0.1 \ {\rm PeV}^2$. This corresponds to a nucleation temperature  $T_* \approx 600 \ {\rm TeV}$.

\section{Conclusions}

We demonstrated that within the framework of the Mini-Split Supersymmetry it is possible for a phase with a reduced symmetry  to develop at PeV-scale temperatures. The resulting finite temperature vacuum structure  can be quite nontrivial,  leading to a strong first order phase transition and/or the production of domain walls in the early Universe. The exact  cosmological evolution depends on the assumptions regarding the reheating temperature. In the cases considered,  a stochastic gravitational wave background at $\sim 1 \ \rm kHz$ frequencies is predicted. In the minimal formulation of the theory,  the  signal is not sufficiently strong to be detected at LIGO or the next generation of  gravitational wave detectors.

Nevertheless, some extensions of the Mini-Split SUSY scenario may provide a much stronger gravitational wave signal. We showed that this is indeed  the case when one introduces 
more scalars
%more families of particles 
 into the theory. It would be interesting to investigate which other modifications of the model  lead to a nontrivial vacuum structure at finite temperature, while providing a measurable gravitational wave signal.

%Although we focused on the MSSM with supersymmetry broken at the PeV scale, our finding can be applied also to Split-MSSM, where supersymmetry is broken at the scale $\sim 10^{11} \ {\rm GeV}$ and beyond. The  first order phase transition signals discussed here would be  inaccessible via
%gravitational wave detectors in the foreseeable future.  However, those phase transitions could produce topological defects like cosmic strings \cite{Gouttenoire:2019kij} or DW  \cite{Saikawa:2017hiv}, which would produce signals detectable in near-future gravitational wave experiments like the Einstein Telescope or Cosmic Explorer. 

Finally, we note that our findings  open the door to searches  for a nontrivial  vacuum structure of  supersymmetric theories at high temperatures. Those efforts are complementary to the recently proposed gravitational wave search for supersymmetry breaking hidden sectors \cite{Craig:2020jfv}.

\subsection*{Acknowledgments} 

We are very grateful to the anonymous {\it Physical Review D} referee for extremely helpful comments regarding the manuscript. We also thank Andrei Angelescu, Peisi Huang and Harikrishnan Ramani  for useful discussions. 
The work of Y.Z. was supported in part by the U.S. Department of Energy under Award
No.~${\rm DE}$-${\rm SC0009959}$. J.H.Y. was supported by the National Natural Science Foundation of China (NSFC) under Grants ${\rm No.}$\,${\rm 12022514}$, ${\rm No.}$\,${\rm 11875003}$, ${\rm No.}$\,${\rm 12047503}$, and the National Key Research and Development Program of China Grant ${\rm No.}$\,${\rm 2020YFC2201501}$. The work of B.S. is supported in part by the NSF grant ${\rm PHY}$-${\rm 2014075}$.

\bibliography{PeV}

\end{document}